\documentclass[journal]{IEEEtran}
\usepackage{amsfonts}
\IEEEoverridecommandlockouts

\ifCLASSINFOpdf
\else
\fi
\usepackage{epstopdf}
\usepackage{multirow}
\usepackage{cite}
\usepackage{stfloats}
\usepackage{color}
\usepackage{epsfig}
\usepackage{graphicx}
\usepackage{psfig}
\usepackage{epsf}
\usepackage{slashbox}
\usepackage{float}
\usepackage{amssymb }
\usepackage[cmex10]{amsmath}
\usepackage{algorithm} 
\usepackage{algorithmic}
\usepackage{booktabs}
\begin{document}

\title{Energy Efficiency Maximization for SWIPT Enabled Two-Way DF Relaying}
\author{Liqin Shi, Yinghui Ye, Rose Qingyang Hu,~\IEEEmembership{Senior Member,~IEEE,} and Hailin Zhang,~\IEEEmembership{Member,~IEEE}\vspace*{-10pt}
\thanks{Personal use of this material is permitted. However, permission to use this
material for any other purposes must be obtained from the IEEE by sending
a request to pubs-permissions@ieee.org.}
\thanks{Liqin Shi, Yinghui Ye and Hailin Zhang  are with the ISN of Xidian University, Xi'an, China.
Rose Qingyang Hu  is with the Department of ECE, Utah State University, USA. The corresponding author is Yinghui Ye (connectyyh@126.com).}
\thanks{
This work was supported by the  scholarship from China Scholarship Council.
}
}
\markboth{IEEE Signal Processing Letters, No. XX, MONTH YY, YEAR 2019}
{Shi\MakeLowercase{\textit{et al.}}: Outage Performance Optimization for SWIPT Enabled Two-Way DF Relaying}
\maketitle

\begin{abstract}
 This paper focuses on the design of an optimal resource allocation scheme to maximize the energy efficiency (EE) in   a simultaneous wireless information and power transfer (SWIPT) enabled two-way decode-and-forward (DF) relay network under a non-linear energy harvesting model. In particular, we  formulate an optimization problem by jointly optimizing the transmit powers of  two source nodes,  the power-splitting (PS) ratios of the relay, and the time for the source-relay transmission, under multiple constraints including  the transmit power constraints at sources and the minimum rate requirement. Although the formulated problem is non-convex, an iterative algorithm is developed to obtain the optimal resource allocation. Simulation results verify the proposed algorithm and show that the designed resource allocation scheme is superior to other benchmark schemes in terms of EE.
\end{abstract}

\IEEEpeerreviewmaketitle
\vspace*{-10pt}
\section{Introduction}
\IEEEPARstart{S}{imultaneous} wireless information
and power transfer {\color{black}(SWIPT) has been recognized as a promising approach to solving the energy scarcity problem in energy-constrained wireless networks, e.g., the Internet of Things (IoT) networks\cite{6957150,7744827,ShiTvt,7054723,7931537}. }Generally, there are two practical schemes to  facilitate SWIPT, namely, time-switching (TS), and power-splitting (PS) schemes.
For the TS scheme, the receiver switches the decoding information and the
harvesting energy in time domain while for the PS scheme, the received signal is split into two different power streams in power domain, one for energy harvesting (EH) and the other for information decoding.

Moreover, wireless relaying has been promoted as an effective solution to improving system energy efficiency (EE), to extending coverage, and to enhancing capacity, particularly in the deep fading of wireless propagations. However,  the limited energy supply in the relay node may discourage  the relay from engaging information transmission. Inspired by this, SWIPT enabled wireless relaying was proposed and devoted to solving this problem \cite{8337780,8355777}.
For the SWIPT enabled one-way relay network, the authors in \cite{8337780} proposed an optimal transmission scheme of {\color{black}joint time allocation and PS to maximize the system capacity under the decode-and-forward (DF) protocol.} In \cite{8355777}, the authors studied the optimal PS/TS ratio for the one-way amplify-and-forward (AF) relay network with a non-linear EH model. Since two-way relaying achieves a higher spectral efficiency, SWIPT enabled two-way relaying has attracted extensive interests \cite{7037438,8361446}.
In \cite{7037438}, the authors studied the achievable throughput of SWIPT enabled AF two-way relaying under three wireless power transfer policies. Authors in \cite{8361446} proposed an optimal asymmetric PS scheme for the SWIPT enabled product two-way AF relaying network to minimize the system outage probability.
Recall that EE has become an emerging important metric in future wireless networks,
the EE was maximized by optimizing both precoding matrices and the PS ratio for SWIPT enabled multiple-input-multiple-output (MIMO) two-way AF networks \cite{7876801,8325514}. By utilizing the statistical channel state information (CSI), the authors in \cite{7858680} studied the optimal power allocation at sources to maximize the EE of SWIPT enabled two-way AF networks. Further, the EE fairness was investigated for multi-pair wireless-powered AF relaying systems in \cite{8474356}.
To the best of our knowledge, there has been  no open work on the EE maximization for SWIPT enabled two-way DF relaying.

In this paper, we focus on the EE maximization for SWIPT enabled two-way DF relay networks under a practical non-linear EH model instead of the linear one adopted in \cite{7876801,8325514,7858680,8474356}. The optimization problem is formulated by jointly optimizing the transmit powers for the two source nodes, the PS ratios at the relay, and the time used for the source-relay transmission. The non-linear EH model makes our formulated problem much more  challenging  than that based on the linear EH model. Towards that end, we propose an iterative algorithm to obtain the optimal solution. Simulation results demonstrate the superiority of the proposed resource allocation scheme in terms of EE.

\vspace*{-15pt}
\section{System model}
\begin{figure}
  \centering
  \includegraphics[width=0.35\textwidth]{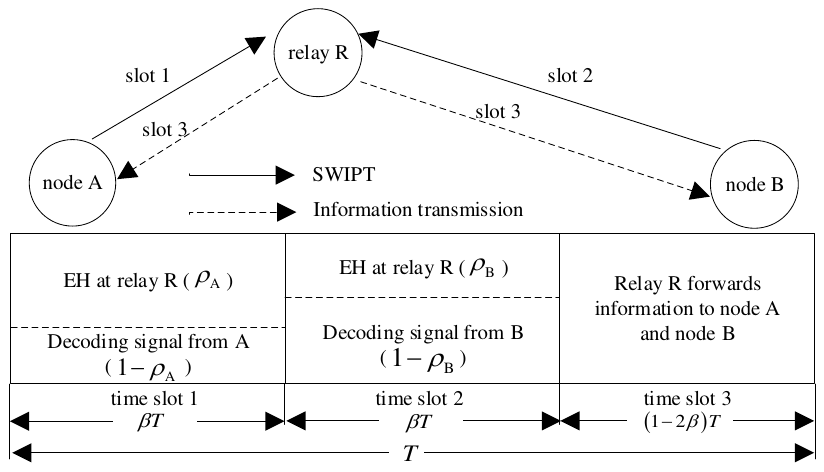}\\
  \caption{{\color{black}System model of the SWIPT enabled two-way DF relaying network.}}\label{fig0}
  \vspace*{-20pt}
\end{figure}
{\color{black}As shown in Fig. 1,
we consider a SWIPT enabled two-way DF relaying network, where both the time division broadcast (TDBC) protocol\footnote{{\color{black}Since the low complexity of hardware is very vital to energy-constrained networks and the operational complexity of the TDBC protocol is lower than that of the Multiple Access Broadcast (MABC) protocol, this work considers the TDBC protocol to realize two-way relaying.}}  and the \lq\lq harvest-then-forward\rq\rq $ $ scheme are employed.}
Two destination nodes $\rm{A}$ and $\rm{B}$ exchange information via an energy-constrained relay node $\rm{R}$.
Assume that no direct link exists between nodes $\rm{A}$ and $\rm{B}$ due to severe path loss and/or shadowing
and each node works in the half-duplex mode with a single antenna equipped. Let $h_{i}$ ($i = \rm{A} \;\rm{or}\; \rm{B}$) denote the channel coefficient between nodes $i$ and $\rm{R}$. Each channel is assumed to undergo independent identically distributed (i.i.d) quasi-static block fading that is reciprocal in two directions.
{\color{black}Here, we assume that the perfect CSI{\footnote{{\color{black}The assumption of perfect CSI can obtain the upper bound of performance for our considered network. Thus, there are also many state-of-art works \cite{8361446,7876801,8325514,8474356}, which are studied based on the assumption of perfect CSI. This motivates us to consider the assumption of perfect CSI.}}} between the relay and two source nodes is available. Specifically,  each source node needs to know the CSI between the source and the relay. The CSI is used to perform successive interference cancellation (SIC) at the source node. The relay needs to know the CSI between the relay and the two source nodes to determine the optimal resource allocation policy to maximize the EE of the whole system. Then the optimal resource allocation policy is known at each source node via the feedback before data transmission in each transmission block. Note that the CSI can be obtained following \cite{1603719} and \cite{8633928}.}

Let $T$ be the duration of the entire transmission block, which  can be divided into three time slots  based on time proportion factor $\beta\in(0,0.5)${\footnote{{\color{black}In this work, it is assumed that the transmission time length of the $\rm{A}$-$\rm{R}$ link equals that of the $\rm{B}$-$\rm{R}$ link, the same as in \cite{8361446,8633928,7782410}. It is worth pointing
out that allowing different time lengths for the $\rm{A}$-$\rm{R}$ and $\rm{B}$-$\rm{R}$ links  can further improve
the performance of the SWIPT enabled two-way DF relaying network, which
is beyond the scope of this paper and will be considered in our future work.}}}. At the first or the second time slot of duration
$\beta T$, $\rm{A}$ or $\rm{B}$ transmits the signal $s_{\rm{A}}$ or $s_{\rm{B}}$ to $\rm{R}$ with the transmit power $P_{\rm{A}}$ or $P_{\rm{B}}$ and the received signal from $i$ ($i = \rm{A} \;\rm{or}\; \rm{B}$) at $\rm{R}$ is given by ${y_{ i\rm{R}}} = {h_{ i}}\sqrt {{P_{i}}} {s_{ i}} + {n_{ \rm{R}}}$,
where {\color{black}${n_{\rm{R}}} \sim {\rm{{\cal C}{\cal N}}}\left( {0,W\sigma^2} \right)$ is the additive white Gaussian noise (AWGN); $W$ denotes the bandwidth of the system, and $\sigma^2$ is  the noise power spectral density.}
After receiving signal from $i$ ($i = \rm{A} \;\rm{or}\; \rm{B}$), $\rm{R}$ splits it into two parts with ratio $\rho_{i}$, where one part is used for energy harvesting and the other part  is used for information processing.

For the energy harvesting, we employ a more practical piecewise linear EH model  \cite{ShiICC, 8355777} and the harvested power $P^{i}_{\rm{H}}$ from $i$ can be computed as
\begin{align}\label{3}
{P^{i}_{\rm{H}}} = \left\{ {\begin{array}{*{20}{c}}
{{\rm{0}},\;\;\;\;\;\;\;\;\;\;\;\;{P^{i}_{\rm{RF}}} \in \left[ {P_{\rm{th}}^0,P_{\rm{th}}^{1}} \right]};\\
\!\!\!\!{{a_j}{P^{i}_{\rm{RF}}} + {b_j},{P^{i}_{\rm{RF}}} \in \left[ {P_{\rm{th}}^j,P_{\rm{th}}^{j + 1}} \right]}\\
\;\;{{P_{\rm{m}}},\;\;\;\;\;\;\;\;\;{P^{i}_{\rm{RF}}} \in \left[ {P_{\rm{th}}^N,P_{\rm{th}}^{N+1}} \right]},
\end{array}} \right.\!\!\!,j = 1,\cdots,N\!-\!\!1;
\end{align}
where ${P^{i}_{\rm{RF}}} = {\rho _i}{P_i}|{h_i}{|^2}$ is the received power from $i$ at $\rm{R}$; ${P_{\rm{th}}} = \{ P_{\rm{th}}^j|0 \le j \le N+1\} $ with $P_{\rm{th}}^0=0$ and $P_{\rm{th}}^{N+1}=+\infty$ are the thresholds on $P^{i}_{\rm{RF}}$ for $N+1$ linear segments;
$a_{j}$ and $b_{j}$ are the scope and the intercept for the linear function in the $j$-th $(j\in\{0,\cdots,N\})$ segment, respectively, and $P_{\rm{m}}$ denotes the maximum harvestable power when the circuit is saturated.
According to \eqref{3}, we have $a_{0}=b_{0}=a_{N}=0$ and $b_{N}=P_{\rm{m}}$.

{\color{black}For the information processing, the received achievable rate at $\rm{R}$ from node $i$ is given by $\tau_{i\rm{R}}=\beta TW\log_{2}\left(1+\frac{{{P_i}|{h_i}{|^2}\left( {1 - {\rho _i}} \right)}}{{ W\sigma^2}}\right)$.}
During the first two time slots, the total harvested energy is calculated as
\begin{align}\label{R1}\notag
E_{\rm{total}}&=\beta T (P ^{\rm{A}}_{\rm{H}}+P^{\rm{B}}_{\rm{H}})\\
&=\!\beta T\left(\! {{a_j}{{\rho _{\rm{A}}}{P_{\rm{A}}}|{h_{\rm{A}}}{|^2}} \!+\! {a_k} {{\rho _{\rm{B}}}{P_{\rm{B}}}|{h_{\rm{B}}}{|^2}} \!+\! {b_k} \!+\! {b_j}} \!\right),
\end{align}
where $j$ or $k$ ($j,k \in \{0,\cdots,N\}$) denotes the segment that ${P^{\rm{A}}_{\rm{RF}}}$ or ${P^{\rm{B}}_{\rm{RF}}}$ belongs to.

In the remaining block time of duration $(1-2\beta)T$, $\rm{R}$ combines the information from $\rm{A}$ and $\rm{B}$ as $s_{\rm{R}}=\frac{\widetilde{s}_{\rm{A}}+\widetilde{s}_{\rm{B}}}{\sqrt{2}}$, where $\widetilde{s}_{\rm{A}}$ and $\widetilde{s}_{\rm{B}}$ denote the decoded signals for $\rm{A}$ and $\rm{B}$,
and broadcasts the combined signal $s_{\rm{R}}$ with the harvested energy $E_{\rm{total}}$.
Accordingly, the received signal at $i$ is written as
\begin{align}\label{6}
{y_{{\rm{R}}i}} &= {h_i}\sqrt {P_{\rm{R}}} {s_{\rm{R}} + n_{i}}
\overset{\text{(a)}}{=}{h_i}\sqrt {P_{\rm{R}}} {\frac{\widetilde{s}_{\overline{i}}}{\sqrt{2}}} + {\widetilde{n}_{i}},
\end{align}
where $P_{\rm{R}} = \frac{{{E_{\rm{total}}}}}{{\left( {1 - 2\beta } \right)T}}$ is the transmit power at $\rm{R}$; {\color{black}${n_{i}}=\widetilde{n}_{i} \sim {\rm{{\cal C}{\cal N}}}\left( {0,W\sigma^2} \right)$ is the AWGN at $i$};
step (a) follows by using SIC \cite{8361446,7037438};  $\bar i$ denotes the index of the other destination node.
{\color{black}Based on \eqref{6}, the achievable rate at $i$ from $\rm{R}$ is given by $\tau_{{\rm{R}}i}=\left(1-2\beta\right) TW\log_{2}\left(1+\frac{{\beta {K_i}}}{{1{\rm{ - }}2\beta }}\left( {\phi \left( {{\rho _{\rm{A}}},{\rho _{\rm{B}}}} \right) + {K_1}} \right)\right)$,
where $\phi \left( {{\rho _{\rm{A}}},{\rho _{\rm{B}}}} \right) =  {{a_j}{\rho _{\rm{A}}}P_{\rm{A}}|{h_{\rm{A}}}{|^2} + {a_k}{\rho _{\rm{B}}}P_{\rm{B}}|{h_{\rm{B}}}{|^2}} $, ${K_1} = {b_k} + {b_j}$ and ${K_i} = \frac{{|{h_i}{|^2}}}{{2W{\sigma ^2}}}$.}

Accordingly, the achievable rate of the link $\bar i\mathop  \to \limits^{\rm{R}} i$ is $R_{\bar i i}=\min\left(\tau_{\bar i\rm{R}}, \tau_{{\rm{R}}i}\right)$.
Then the total system achievable rate is given by $R_{\rm{AB}}+R_{\rm{BA}}$.
The total system energy consumption can be computed as $\beta T\left( {{P_{\rm{A}} \mathord{\left/
 {\vphantom {P \varepsilon }} \right.
 \kern-\nulldelimiterspace} \varepsilon } +{P_{\rm{B}} \mathord{\left/
 {\vphantom {P \varepsilon }} \right.
 \kern-\nulldelimiterspace} \varepsilon }+ {P_{ct}}} \right) + 2{P_{cr}}\left( {1 - 2\beta } \right)T$, where $\varepsilon$ denotes the power amplifier efficiency
of the sources; $P_{ct}$ is the constant circuit power consumption at nodes $\rm{A}$ and $\rm{B}$ as transmitters and $P_{cr}$ is the  constant circuit power consumed at nodes $\rm{A}$ and $\rm{B}$ as receivers.

\vspace*{-10pt}
\section{Energy Efficiency Maximization}
\subsection{Problem Formulation}
Considering that the non-linearity of the practical energy harvester always makes the optimization problem non-convex and difficult to solve, here we apply the piecewise linear EH model and
the optimization problem to maximize the EE of the system can be formulated as $\mathbf{P}_1$.
\begin{align}\nonumber
\begin{array}{*{20}{l}}
\mathbf{P}_1: {\mathop {\max }\limits_{j,k}\;\;\mathop {{\rm{max}}}\limits_{\left( {P_{\rm{A}},P_{\rm{B}},\beta ,{\rho _{\rm{A}}}{\rm{,}}{\rho _{\rm{B}}}} \right)} \;\; {\rm{ }}\frac{{{R_{{\rm{AB}}}} + {R_{{\rm{BA}}}}}}{{\beta T\left( {{P_{\rm{A}} \mathord{\left/
 {\vphantom {P \varepsilon }} \right.
 \kern-\nulldelimiterspace} \varepsilon } +{P_{\rm{B}} \mathord{\left/
 {\vphantom {P \varepsilon }} \right.
 \kern-\nulldelimiterspace} \varepsilon }+ {P_{ct}}} \right) + 2{P_{cr}}\left( {1 - 2\beta } \right)T }}}\\
\begin{array}{l}
{\rm{s}}.{\rm{t}}.\;:\;{\rm{ }}{\rm{C1}}:{\rm{ }}{R_{{\rm{AB}}}} \ge {R_{\min }};\\
{\rm{       }}\;\;\;\;\;\;\;\;\;\;{\rm{C2}}:{\rm{ }}{R_{{\rm{BA}}}} \ge {R_{\min }};\\
{\rm{       }}\;\;\;\;\;\;\;\;\;\;{\rm{C3}}:{\rm{ 0 < }}P_i \le {P_{\max }},i \in \left\{ {{\rm{A, B}}} \right\};\!\!\\
{\rm{       }}\;\;\;\;\;\;\;\;\;\;{\rm{C4}}:{\rm{ 0 < }}\beta {\rm{ < 0}}{\rm{.5}};\\
{\rm{       }}\;\;\;\;\;\;\;\;\;\;{\rm{C5}}:{\rm{ 0 < }}{\rho _i} < 1,{\rm{ }}i \in \left\{ {{\rm{A, B}}} \right\};\\
{\rm{       }}\;\;\;\;\;\;\;\;\;\;{\rm{C6}}:P_{{\rm{th}}}^{j} \le {\rho _{\rm{A}}}P_{\rm{A}}|{h_{\rm{A}}}{|^2} \le P_{{\rm{th}}}^{j+1},j\in\{ 0,\cdots,N\};\\
{\rm{       }}\;\;\;\;\;\;\;\;\;\;{\rm{C7}}:P_{{\rm{th}}}^{k} \le {\rho _{\rm{B}}}P_{\rm{B}}|{h_{\rm{B}}}{|^2} \le P_{{\rm{th}}}^{k+1},k\in\{ 0,\cdots,N\};
\end{array}
\end{array}
\end{align}
where constrains $\rm{C1}$ and $\rm{C2}$ ensure the minimum required rate $R_{\min}$ for each end-to-end link; $P_{\max}$ is the maximum transmission power at sources;  $\rm{C6}$ and $\rm{C7}$ are the constraints that the energy harvester works in the $j$-th and $k$-th ($j,k\in\left\{0,\cdots,N\right\}$) linear regions, respectively.

In order to solve $\mathbf{P}_1$, there are three main steps as follows. In the first step, we compute the maximum number of segments that ${P^{\rm{A}}_{\rm{RF}}}$ and ${P^{\rm{B}}_{\rm{RF}}}$ may belong to, denoted by $s_{\rm{A}}$ and $s_{\rm{B}}$ ($s_{\rm{A}}$, $s_{\rm{B}}$ $\in\left\{0,\cdots,N\right\}$), respectively. {\color{black}Specifically, $s_{\rm{A}}$ is the maximum number of segments which satisfies $P_{\rm{max}}|h_{\rm{A}}|^2\geq P_{\rm{th}}^{s_{\rm{A}}}$. }
In the second step,  we solve the following optimization problem $\mathbf{P}_2$  for given parameters $j$ and $k$ ($j\in\left\{0,\cdots,s_{\rm{A}}\right\},k\in\left\{0,\cdots,s_{\rm{B}}\right\}$):
\begin{align}\nonumber
\begin{array}{*{20}{l}}
\mathbf{P}_2: {\;\mathop {{\rm{max}}}\limits_{\left( {P_{\rm{A}},P_{\rm{B}},\beta ,{\rho _{\rm{A}}},{\rho _{\rm{B}}}} \right)} \;\;\frac{{{R_{{\rm{AB}}}} + {R_{{\rm{BA}}}}}}{{\beta T\left( {{P_{\rm{A}} \mathord{\left/
 {\vphantom {P \varepsilon }} \right.
 \kern-\nulldelimiterspace} \varepsilon } +{P_{\rm{B}} \mathord{\left/
 {\vphantom {P \varepsilon }} \right.
 \kern-\nulldelimiterspace} \varepsilon }+ {P_{ct}}} \right) + 2{P_{cr}}\left( {1 - 2\beta } \right)T }}}\\
{{\rm{s}}.{\rm{t}}.\;:\;{\rm{C1 - C7}}{\rm{.}}}
\end{array}
\end{align}
{\color{black}Note that for the case with $j+k=0$, the total harvested energy is always $0$. Then the total system achievable rate is $0$ and the constraints $\rm{C1}$ and $\rm{C2}$ can not be guaranteed. In this case, the optimization problem $\mathbf{P}_2$ is infeasible.
For the cases with $j+k\neq 0$, by solving $\mathbf{P}_2$, we can obtain $(s_{\rm{A}}+1)(s_{\rm{B}}+1)-1$ resource allocation policies, denoted by $(P^{j,k}_{\rm{A}},P^{j,k}_{\rm{B}},\beta^{j,k},\rho^{j,k}_{\rm{A}},\rho^{j,k}_{\rm{B}})_{j,k \in \left\{ {0,...,{\rm{N}}} \right\}\backslash j = k = 0}$, and compute the corresponding EE as $(q^{j,k})_{j,k \in \left\{ {0,...,{\rm{N}}} \right\}\backslash j = k = 0}$.
In the third step, we compare these $(s_{\rm{A}}+1)(s_{\rm{B}}+1)-1$ values of EE and find the optimal solution to $\mathbf{P}_1$, denoted by $(P^+_{\rm{A}},P^+_{\rm{B}},\beta^+,\rho_{\rm{A}}^+,\rho_{\rm{B}}^+)$.
Specifically, the optimal solution to $\mathbf{P}_1$ is determined by $\arg {\rm{max}}\left\{ {{{\left( {{q^{j,k}}} \right)}_{j,k \in \left\{ {0,...,{\rm{N}}} \right\}\backslash j = k = 0}}} \right\}$.}
Note that the main difficulty is to solve the non-convex fractional optimization problem $\mathbf{P}_2$ due to the existence of coupling relationship among different optimization variables and the non-convex constraints, i.e., $\rm{C1}, \rm{C2}, \rm{C6}$ and $\rm{C7}$.

\vspace*{-10pt}
\subsection{Solution to  $\mathbf{P}_2$}
To deal with the coupling relationship between $\beta$ and $P_i$ ($i \in \left\{ {{\rm{A, B}}} \right\}$), we let $t=\frac{1-2\beta}{\beta}$ and $\mathbf{P}_2$ can be rewritten as
\begin{align}\nonumber
\begin{array}{l}
\mathbf{P}_3: \mathop {{\rm{max}}}\limits_{\left( {P_{\rm{A}},P_{\rm{B}},t ,{\rho _{\rm{A}}},{\rho _{\rm{B}}}} \right)} \;\;\frac{{R_{{\rm{AB}}}^{\left( 1 \right)} + R_{{\rm{BA}}}^{\left( 1 \right)}}}{{ {{P_{\rm{A}} \mathord{\left/
 {\vphantom {P \varepsilon }} \right.
 \kern-\nulldelimiterspace} \varepsilon }}+{{P_{\rm{B}} \mathord{\left/
 {\vphantom {P \varepsilon }} \right.
 \kern-\nulldelimiterspace} \varepsilon } + {P_{ct}}} + 2{P_{cr}}t}}\\
{\rm{s}}.{\rm{t}}.\;:\;{\rm{C1-1}}:R_{{\rm{AB}}}^{\left( 1 \right)} \ge {R_{\min }}\left(t+2\right);\\
\;\;\;\;\;\;\;\;\;\;{\rm{      C2-1}}:R_{{\rm{BA}}}^{\left( 1 \right)} \ge {R_{\min }}\left(t+2\right);\\
\;\;\;\;\;\;\;\;\;\;{\rm{      C4-1}}:t > {\rm{0}}; C3,C5,C6,C7;
\end{array}
\end{align}
where {\color{black}$R_{{\rm{AB}}}^{\left( 1 \right)} =W\times$\\$\min\!\big( {{{\log }_2}\!\left( {1\! +\!  \frac{{P_{\rm{A}}|{h_{\rm{A}}}{|^2}\left( {1 - {\rho _{\rm{A}}}} \right)}}{{{W\sigma ^2}}}} \right),t{{\log }_2}\!\big( {1 \!+\! \frac{{{K_{\rm{B}}}}}{t}
\big( {\phi \left( {{\rho _{\rm{A}}},{\rho _{\rm{B}}}} \right) \!+\! {K_1}} \!\big)} \!\big)}\! \big)$ and $R_{{\rm{BA}}}^{\left( 1 \right)} =W\times \min ({\log _2}\left( {1 + \frac{{P_{\rm{B}}|{h_{\rm{B}}}{|^2}\left( {1 - {\rho _{\rm{B}}}} \right)}}{{{W\sigma ^2}}}} \right),t{\log _2}(1 + \frac{{{K_{\rm{A}}}}}{t}(\phi \left( {{\rho _{\rm{A}}},{\rho _{\rm{B}}}} \right) + {K_1})))$.}

{\color{black}Let $(P^*_{\rm{A}},P^*_{\rm{B}},t^*,\rho_{\rm{A}}^*,\rho_{\rm{B}}^*)$ denote the optimal solution to $\mathbf{P}_3$. It is worth noting that the optimal solution to $\mathbf{P}_2$ is given by $(P^*_{\rm{A}},P^*_{\rm{B}},\beta^*,\rho_{\rm{A}}^*,\rho_{\rm{B}}^*)$, where $\beta^*=\frac{1}{t^*+1}.$ The maximum EE for given $j$ and $k$, denoted by $q^*$, can be defined as}
\begin{align}\label{b51}\notag
&{q^*} = \frac{{R_{{\rm{AB}}}^{\left( 1 \right)}\left( {{P^*_{\rm{A}}},P^*_{\rm{B}},{t^*},\rho _{\rm{A}}^*,\rho _{\rm{B}}^*} \right) + R_{{\rm{BA}}}^{\left( 1 \right)}\left( {{P^*_{\rm{A}}},P^*_{\rm{B}},{t^*},\rho _{\rm{A}}^*,\rho _{\rm{B}}^*} \right)}}{{{{P^*_{\rm{A}} \mathord{\left/
 {\vphantom {P \varepsilon }} \right.
 \kern-\nulldelimiterspace} \varepsilon }}+{{P^*_{\rm{B}} \mathord{\left/
 {\vphantom {P \varepsilon }} \right.
 \kern-\nulldelimiterspace} \varepsilon } + {P_{ct}}} + 2{P_{cr}}t^*}}\\
&=\begin{array}{l}
\mathop {{\rm{max}}}\limits_{\left( {P_{\rm{A}},P_{\rm{B}},t,{\rho _{\rm{A}}},{\rho _{\rm{B}}}} \right)} \;\;q\left( {P_{\rm{A}},P_{\rm{B}},t,{\rho _{\rm{A}}},{\rho _{\rm{B}}}} \right)\\
{\rm{s}}.{\rm{t}}.\;:\;{\rm{C1 - 1, C2 - 1, C3, C4 - 1, C5 - C7}},
\end{array}
\end{align}
where $q=\frac{{R_{{\rm{AB}}}^{\left( 1 \right)}\left( {P_{\rm{A}},P_{\rm{B}},t,{\rho _{\rm{A}}},{\rho _{\rm{B}}}} \right) + R_{{\rm{BA}}}^{\left( 1 \right)}\left( {P_{\rm{A}},P_{\rm{B}},t,{\rho _{\rm{A}}},{\rho _{\rm{B}}}} \right)}}{{{{P_{\rm{A}} \mathord{\left/
 {\vphantom {P \varepsilon }} \right.
 \kern-\nulldelimiterspace} \varepsilon }}+{{P_{\rm{B}} \mathord{\left/
 {\vphantom {P \varepsilon }} \right.
 \kern-\nulldelimiterspace} \varepsilon } + {P_{ct}}} + 2{P_{cr}}t}}$.

According to the Dinkelbach's method \cite{6294504}, the maximum EE $q^*$ is achieved if and only if the following equation is satisfied.
\begin{align}\label{b7}
\begin{array}{l}
\mathop {{\rm{max}}}\limits_{\left( {P_{\rm{A}},P_{\rm{B}},t,{\rho _{\rm{A}}},{\rho _{\rm{B}}}} \right)} \!\!\!\!R_{{\rm{AB}}}^{\left( 1 \right)} \!+\! R_{{\rm{BA}}}^{\left( 1 \right)} \!-\! {q^*}\!\left(\frac{{{P_{\rm{A}}} \!+\! {P_{\rm{B}}}}}{\varepsilon }\! +\!{P_{ct}}\!+\! \!2{P_{cr}}t\right)\!\\
 = R_{{\rm{AB}}}^{\left( 1 \right)*} \!+\! R_{{\rm{BA}}}^{\left( 1 \right)*} \!-\! {q^*}\left(\frac{{{P^*_{\rm{A}}} + {P^*_{\rm{B}}}}}{\varepsilon }\!+\! {P_{ct}}+\! 2{P_{cr}}t^*\right) = 0.
\end{array}
\end{align}

{\color{black}Thus, the problem in $\mathbf{P}_3$ can be transformed by solving a parametric
problem $\mathbf{P}_4$.
\begin{align}\nonumber
\begin{array}{*{20}{l}}
\mathbf{P}_4:{\;\mathop {{\rm{max}}}\limits_{\left( {P_{\rm{A}},P_{\rm{B}},t,{\rho _{\rm{A}}},{\rho _{\rm{B}}}} \right)} \!\!\!\!R_{{\rm{AB}}}^{\left( 1 \right)}\! +\! R_{{\rm{BA}}}^{\left( 1 \right)} \!-\! q\left(\frac{{{P_{\rm{A}}} + {P_{\rm{B}}}}}{\varepsilon }\!+\! {P_{ct}} \!+\! 2{P_{cr}}t\right)}\\
{{\rm{s}}.{\rm{t}}.\;:\;{\rm{C1-1,C2-1,C3,C4-1,C5,C6,C7}},}\!\!\!\!\!\!\!\!\!\!\!\!\!\!\!\!\!\!
\end{array}
\end{align}
where $q$ is a given parameter.}

For the optimization problem $\mathbf{P}_4$, it is still non-convex due to the coupling between $P_i$ and $\rho_i$ ($i\in\{\rm{A}, \rm{B}\}$) and non-convex constraints. We introduce four auxiliary variables into the problem. Specifically,
we let $r_{\rm{A}}=R_{{\rm{AB}}}^{\left( 1 \right)}$, $r_{\rm{B}}=R_{{\rm{BA}}}^{\left( 1 \right)}$ and $x_{i}=P_i\rho_{i}, i\in\{\rm{A},\rm{B}\}$. Then the optimization problem  $\mathbf{P}_4$ can be equivalently expressed as $\mathbf{P}_5$.
\begin{align} \nonumber
\begin{array}{*{20}{l}}
\mathbf{P}_5:{\;\mathop {{\rm{max}}}\limits_{\left( {P_{\rm{A}},P_{\rm{B}},t,{x_{\rm{A}}},{x_{\rm{B}}},{r_{\rm{A}}},{r_{\rm{B}}}} \right)} \!\!\!{r_{\rm{A}}} + {r_{\rm{B}}} - q\left(\frac{{{P_{\rm{A}}} + {P_{\rm{B}}}}}{\varepsilon }+ {P_{ct}} + 2{P_{cr}}t\right)}\\
{\rm{s}}.{\rm{t}}.\;:\;{\rm{C1-2:}}{r_{\rm{A}}} \ge {R_{\min }}\left(t+2\right),\\
\;\;\;\;\;\;\;\;\;\;{\rm{C2-2:}}{r_{\rm{B}}} \ge {R_{\min }}\left(t+2\right),{\rm{C3}},{\rm{C4-1}},\\
\;\;\;\;\;\;\;\;\;\;{\rm{C5-1:}}0 < {x_i} < P_i,i \in \left\{ {{\rm{A, B}}} \right\},\\
\;\;\;\;\;\;\;\;\;\;{\rm{C6-1:}}P_{{\rm{th}}}^{j} \le {x_{\rm{A}}}|{h_{\rm{A}}}{|^2} \le P_{{\rm{th}}}^{j+1},\\
\;\;\;\;\;\;\;\;\;\;{\rm{C7-1:}}P_{{\rm{th}}}^{k} \le {x_{\rm{B}}}|{h_{\rm{B}}}{|^2} \le P_{{\rm{th}}}^{k+1},\\
\;\;\;\;\;\;\;\;\;\;{\rm{C8}}:{\color{black}{W\log _2}\left( {1 + \frac{{|{h_i}{|^2}\left( {P_i - {x_i}} \right)}}{{{W\sigma ^2}}}} \right) \ge {r_i},i \in \left\{ {{\rm{A, B}}} \right\}};\\
\;\;\;\;\;\;\;\;\;\;{\rm{C9}}:{\color{black}Wt{\log _2}\left( {1 + \frac{{{K_{\bar i}}}}{t}{\left( {\phi_1 + {K_1}} \right)}} \right) \ge {r_i},\bar i,i \in \left\{ {{\rm{A, B}}} \right\}};
\end{array}
\end{align}
where $\phi_1= {{a_j}{ x_{\rm{A}}}|{h_{\rm{A}}}{|^2} + {a_k}{x_{\rm{B}}}|{h_{\rm{B}}}{|^2}}$.

\textbf{Proposition 1}: The optimization problem in $\mathbf{P}_5$ is convex. 

\emph{Proof:} See the Appendix. \hfill {$\blacksquare $}

\vspace*{-10pt}
\subsection{Algorithm}
In order to solve $\mathbf{P}_2$, we develop a Dinkelbach-based iterative algorithm as shown in Algorithm 1. In Algorithm 1, we solve the optimization problem $\mathbf{P}_5$ with a given $q$ in each iteration and obtain the optimal solution, denoted by $\left( {P'_{\rm{A}},P'_{\rm{B}},t',{x'_{\rm{A}}},{x'_{\rm{B}}},{r'_{\rm{A}}},{r'_{\rm{B}}}} \right)$. Given an error tolerance $\epsilon$, when ${r'_{\rm{A}}} + {r'_{\rm{B}}} - q\left(\frac{{{P'_{\rm{A}}} + {P'_{\rm{B}}}}}{\varepsilon }\!+\! {P_{ct}} \!+\! 2{P_{cr}}t'\right)<\epsilon$ is satisfied, the solution to $\mathbf{P}_2$ can be obtained.
{\color{black}Using Algorithm 1, the optimal solution to $\mathbf{P}_1$ can be obtained based on the three steps mentioned in Section III-A.
}

{\color{black}
Assume that the interior point method is used to obtain the optimal solution to $\mathbf{P}_5$ and that the number of iterations for Algorithm 1 is $N_u$. Based on \cite{Cvex}, the computational complexity for Algorithm 1 can be computed as $N_uO(\sqrt{m_1}\log(m_1))$, where $m_1$ denotes the number of the inequality constraints for $\mathbf{P}_5$. Then the computational complexity to solve $\mathbf{P}_1$ can be computed as $\left[\left(s_{\rm{A}}+1\right)\left(s_{\rm{B}}+1\right)-1\right]N_uO(\sqrt{m_1}\log(m_1))$.}
\begin{figure*}[!t]
\normalsize
\setcounter{equation}{5}
\begin{align}
\frac{{{\partial ^2}{f_2}(t,{x_{\rm{A}}},{x_{\rm{B}}})}}{{\partial {{(t,{x_{\rm{A}}},{x_{\rm{B}}})}^2}}} = \left[ {\begin{array}{*{20}{c}}
{\frac{{ - {{\left( {{c_{\bar i{\rm{A}}}}{x_{\rm{A}}} + {c_{\bar i{\rm{B}}}}{x_{\rm{B}}} + {c_{\bar i{\rm{1}}}}} \right)}^2}}}{{t{{\left( {{c_{\bar i{\rm{A}}}}{x_{\rm{A}}} + {c_{\bar i{\rm{B}}}}{x_{\rm{B}}} + {c_{\bar i{\rm{1}}}} + t} \right)}^2}\ln2}}}&{\frac{{{c_{\bar i{\rm{A}}}}\left( {{c_{\bar i{\rm{A}}}}{x_{\rm{A}}} + {c_{\bar i{\rm{B}}}}{x_{\rm{B}}} + {c_{\bar i{\rm{1}}}}} \right)}}{{{{\left( {{c_{\bar i{\rm{A}}}}{x_{\rm{A}}} + {c_{\bar i{\rm{B}}}}{x_{\rm{B}}} + {c_{\bar i{\rm{1}}}} + t} \right)}^2}\ln2}}}&{\frac{{{c_{\bar i{\rm{B}}}}\left( {{c_{\bar i{\rm{A}}}}{x_{\rm{A}}} + {c_{\bar i{\rm{B}}}}{x_{\rm{B}}} + {c_{\bar i{\rm{1}}}}} \right)}}{{{{\left( {{c_{\bar i{\rm{A}}}}{x_{\rm{A}}} + {c_{\bar i{\rm{B}}}}{x_{\rm{B}}} + {c_{\bar i{\rm{1}}}} + t} \right)}^2}\ln2}}}\\
{\frac{{{c_{\bar i{\rm{A}}}}\left( {{c_{\bar i{\rm{A}}}}{x_{\rm{A}}} + {c_{\bar i{\rm{B}}}}{x_{\rm{B}}} + {c_{\bar i{\rm{1}}}}} \right)}}{{{{\left( {{c_{\bar i{\rm{A}}}}{x_{\rm{A}}} + {c_{\bar i{\rm{B}}}}{x_{\rm{B}}} + {c_{\bar i{\rm{1}}}} + t} \right)}^2}\ln2}}}&{\frac{{ - c_{\bar i{\rm{A}}}^2t}}{{{{\left( {{c_{\bar i{\rm{A}}}}{x_{\rm{A}}} + {c_{\bar i{\rm{B}}}}{x_{\rm{B}}} + {c_{\bar i{\rm{1}}}} + t} \right)}^2}\ln2}}}&{\frac{{ - {c_{\bar i{\rm{A}}}}{c_{\bar i{\rm{B}}}}t}}{{{{\left( {{c_{\bar i{\rm{A}}}}{x_{\rm{A}}} + {c_{\bar i{\rm{B}}}}{x_{\rm{B}}} + {c_{\bar i{\rm{1}}}} + t} \right)}^2}\ln2}}}\\
{\frac{{{c_{\bar i{\rm{B}}}}\left( {{c_{\bar i{\rm{A}}}}{x_{\rm{A}}} + {c_{\bar i{\rm{B}}}}{x_{\rm{B}}} + {c_{\bar i{\rm{1}}}}} \right)}}{{{{\left( {{c_{\bar i{\rm{A}}}}{x_{\rm{A}}} + {c_{\bar i{\rm{B}}}}{x_{\rm{B}}} + {c_{\bar i{\rm{1}}}} + t} \right)}^2}\ln2}}}&{\frac{{ - {c_{\bar i{\rm{A}}}}{c_{\bar i{\rm{B}}}}t}}{{{{\left( {{c_{\bar i{\rm{A}}}}{x_{\rm{A}}} + {c_{\bar i{\rm{B}}}}{x_{\rm{B}}} + {c_{\bar i{\rm{1}}}} + t} \right)}^2}\ln2}}}&{\frac{{ - c_{\bar i{\rm{B}}}^2t}}{{{{\left( {{c_{\bar i{\rm{A}}}}{x_{\rm{A}}} + {c_{\bar i{\rm{B}}}}{x_{\rm{B}}} + {c_{\bar i{\rm{1}}}} + t} \right)}^2}\ln2}}}
\end{array}} \right]
\end{align}
\hrulefill
\setcounter{equation}{6}
\vspace*{-10pt}
\end{figure*}

\begin{algorithm}
\caption{Dinkelbach-based Iterative Algorithm for $\mathbf{P}_2$}
\label{alg:A}
\begin{algorithmic}[1]
\STATE {Initialize the maximum iterations $L_{\max}$ and the maximum error tolerance $\epsilon$.}
\STATE Set the maximum EE $q=0$ and iteration index $l=0$.
\REPEAT
\STATE Solve $\mathbf{P}_5$ with a given $q$ by using CVX and obtain the optimal solution $\left( {P'_{\rm{A}},P'_{\rm{B}},t',{x'_{\rm{A}}},{x'_{\rm{B}}},{r'_{\rm{A}}},{r'_{\rm{B}}}} \right)$;
\IF {${r'_{\rm{A}}} + {r'_{\rm{B}}} - q\left(\frac{{{P'_{\rm{A}}} + {P'_{\rm{B}}}}}{\varepsilon }\!+\! {P_{ct}} \!+\! 2{P_{cr}}t'\right)<\epsilon$}
\STATE Set $P^{\ast}_{\rm{A}}=P'_{\rm{A}}, P^{\ast}_{\rm{B}}=P'_{\rm{B}},\beta^*=\frac{1}{t'+2}, \rho^*_{\rm{A}}=\frac{{x'_{\rm{A}}}}{P'_{\rm{A}}}, \rho^*_{\rm{B}}=\frac{{x'_{\rm{B}}}}{P'_{\rm{B}}}, q^{\ast}=\frac{{r'_{\rm{A}}}+{r'_{\rm{B}}}}{{{{P'_{\rm{A}} \mathord{\left/
 {\vphantom {P \varepsilon }} \right.
 \kern-\nulldelimiterspace} \varepsilon }}+{{P'_{\rm{B}} \mathord{\left/
 {\vphantom {P \varepsilon }} \right.
 \kern-\nulldelimiterspace} \varepsilon } + {P_{ct}}} + 2{P_{cr}}t'}}, \textrm{Flag}=1$ and return
\ELSE
\STATE Set $q=\frac{{r'_{\rm{A}}}+{r'_{\rm{B}}}}{{{{P'_{\rm{A}} \mathord{\left/
 {\vphantom {P \varepsilon }} \right.
 \kern-\nulldelimiterspace} \varepsilon }}+{{P'_{\rm{B}} \mathord{\left/
 {\vphantom {P \varepsilon }} \right.
 \kern-\nulldelimiterspace} \varepsilon } + {P_{ct}}} + 2{P_{cr}}t'}}, l=l+1, \textrm{Flag}=0$
\ENDIF
\UNTIL{$\textrm{Flag}=1$ or $l=L_{\max}$}
\end{algorithmic}
\vspace*{-5pt}
\end{algorithm}

\vspace*{-10pt}
\section{Simulations}
In this section, we evaluate the effectiveness of our proposed EE maximization framework via computer simulations.
In particular, the distance-dependent pathloss model $|g_i|^2d_i^{-\alpha}$ ($i\in\{\rm{A},\rm{B}\}$) is adopted, where $g_i$ is the $i$-$\rm{R}$ channel coefficient, $d_i$ is the distance between node $i$ and the relay and $\alpha$ denotes the pathloss exponent.
According to \cite{7831382,8325514},
the simulation parameters are set as follows: ${d_{{\rm{A}}}} = 5$m, $d_{\rm{B}}=15$m, $P_{ct}=P_{cr}=10$dBm, $P_{\max}=30$dBm, $\alpha=3$, {\color{black}$W=10$ kHz and $\sigma^{2}  = -120$ $\rm{dBm}/$Hz}.
The power amplifier efficiency is set to be $0.35$.
{\color{black}The minimum required rate is set as $R_{\min}=30$ kbps.} We employ the piecewise linear EH model with $N=4$, where $P_{\rm{th}}=[0,10,57.68,230.06,1000,+\infty]$ uW, $\{a_{k}\}^{4}_{0}=[0,0.3899,0.6967,0.1427,0]$ and $\{b_{k}\}^{4}_{0}=[0,-1.6613,-19.1737,108.2778,250]$ uW \cite{ShiICC}.

Fig. 2 demonstrates the convergence of the proposed Algorithm 1
under different sets of $j$ and $k$. We set $|g_{\rm{A}}|^2=1.0571$ and $|g_{\rm{B}}|^2=1.4131$. It can be observed that with any given $j$ and $k$, the EE always converges to the optimal value within a limited number
of iterations, which indicates that our proposed algorithm is computationally efficient.

Fig. 3 shows the average EE versus the maximum transmit power $P_{\max}$. {\color{black}To verify the effectiveness of the proposed scheme, we compare it with four other schemes under the same constraints. These four schemes are  the equal transmit power allocation scheme (denoted as \lq\lq Equal transmit power $(P_{\rm{A}}=P_{\rm{B}})$\rq\rq), the equal PS ratio scheme (denoted as \lq\lq Equal PS ratio $(\rho_{\rm{A}}=\rho_{\rm{B}})$\rq\rq),  the equal transmit power allocation and PS ratio
scheme (denoted as \lq\lq Equal transmit power and PS ratio $(P_{\rm{A}}=P_{\rm{B}}, \rho_{\rm{A}}=\rho_{\rm{B}})$\rq\rq),  and the throughput maximization scheme.} It can be observed that the throughput maximization scheme achieves a much lower EE than the proposed scheme due to the fact that the optimal resource allocation for SE maximization is not energy efficient, which illustrates the importance  of considering the EE. Another observation is that our proposed scheme can achieve the highest EE among these schemes since the proposed scheme provides more flexibility to utilize the resource efficiently.
\begin{figure}
  \centering
  \includegraphics[width=0.35\textwidth]{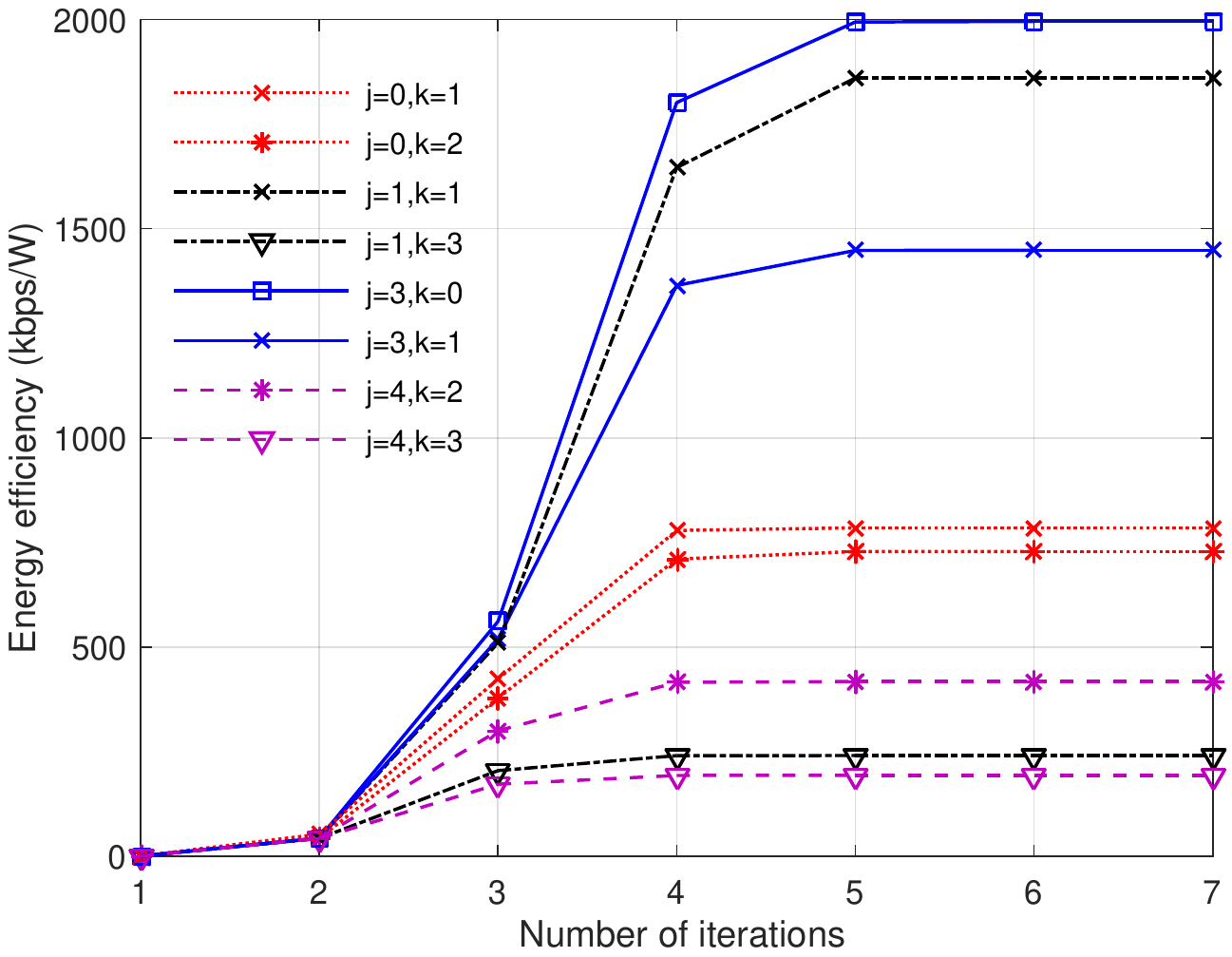}\\
  \caption{{\color{black}Convergence of the proposed Algorithm 1 under different $j$ and $k$.}}\label{fig1}
  \vspace*{-5pt}
\end{figure}
\begin{figure}
  \centering
  \includegraphics[width=0.35\textwidth]{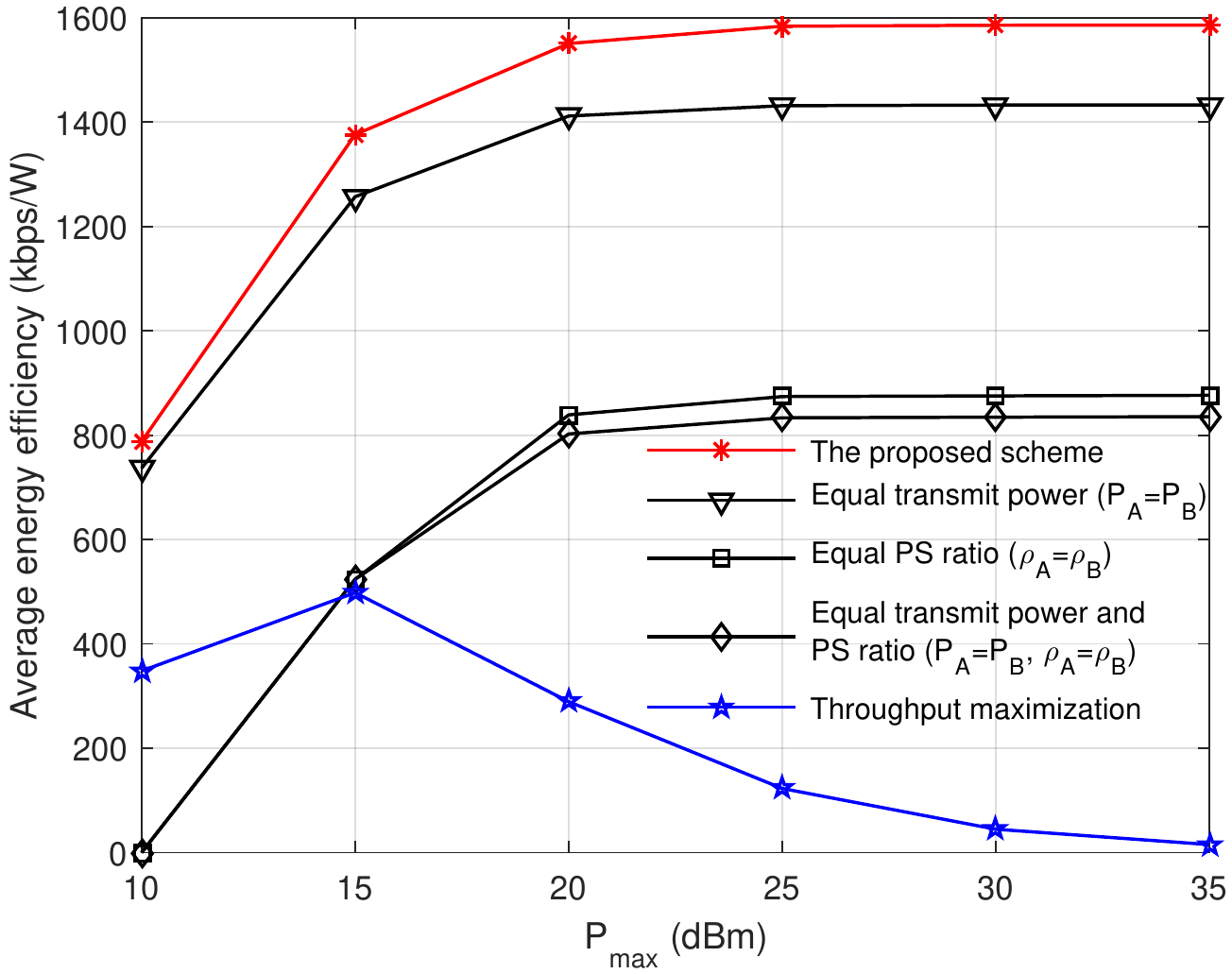}\\
  \caption{{\color{black}Average EE versus the maximum transmit power $P_{\max}$.}}\label{fig2}
  \vspace*{-20pt}
\end{figure}

\vspace*{-10pt}
\section{Conclusions}
In this paper, we have studied the EE optimization for  SWIPT enabled two-way DF relay networks, where a non-linear energy
harvester is equipped at the relay. We have formulated the EE maximization problem by jointly optimizing the transmit powers for two source nodes, the PS ratios at the relay, and the time for the source-relay transmission and proposed an iterative algorithm to achieve the optimal solution and the maximum EE. Simulation results illustrate the superiority of our proposed resource allocation scheme in terms of the EE.

\vspace*{-15pt}
\section*{Appendix}
Since the objective function of $\mathbf{P}_5$ is a linear function with respect to $t$, $r_{i}$ and $P_i$ $i\in\{\rm{A}, \rm{B}\}$, and $\rm{C1-2}, \rm{C2-2}, \rm{C3}, \rm{C4-1}, \rm{C5-1}, \rm{C6-1}$ and $\rm{C7-1}$ are linear constraints, whether $\mathbf{P}_5$ is convex or not depends on constraints $\rm{C8}$ and $\rm{C9}$. {\color{black}That is, if both functions $f_1(P_i,x_i)={\log _2}\left( {1 + \frac{{|{h_i}{|^2}\left( {P_i - {x_i}} \right)}}{{{W\sigma ^2}}}} \right)$ and $f_2(t,x_{\rm{A}},x_{\rm{B}})=t{\log _2}\left( {1 + \frac{{{K_{\bar i}}}}{t}{\left( {{{a_j}{ x_{\rm{A}}}|{h_{\rm{A}}}{|^2} + {a_k}{x_{\rm{B}}}|{h_{\rm{B}}}{|^2}} + {K_1}} \right)}} \right)$, ($i, \bar i=\rm{A}$ or $\rm{B}$), are concave, then $\mathbf{P}_5$ is convex.}
Taking the second-order derivative of $f_1(P_i,x_i)$, the Hessian matrix is given by
$\frac{{{\partial ^2}{f_1}\left( {P_i,{x_i}} \right)}}{{\partial {{\left( {P_i,{x_i}} \right)}^2}}} = \left[ {\begin{array}{*{20}{c}}
{\frac{{ - a_i^2}}{{{{\left( {1 + {a_i}\left( {P_i - {x_i}} \right)} \right)}^2}\ln2}}}&{\frac{{a_i^2}}{{{{\left( {1 + {a_i}\left( {P_i - {x_i}} \right)} \right)}^2}\ln2}}}\\
{\frac{{a_i^2}}{{{{\left( {1 + {a_i}\left( {P_i - {x_i}} \right)} \right)}^2}\ln2}}}&{\frac{{ - a_i^2}}{{{{\left( {1 + {a_i}\left( {P_i - {x_i}} \right)} \right)}^2}\ln2}}}
\end{array}} \right]\leq 0$, where $a_i=2K_i, i\in\{\rm{A}, \rm{B}\}$. Thus, $f_1(P_i,x_i)$ is concave. As for $f_2(t,x_{\rm{A}},x_{\rm{B}})$, the Hessian matrix is given by (6) at the top of this page, where $c_{\bar i{\rm{A}}}=K_{\bar i}a_j|{h_{\rm{A}}}{|^2}$, $c_{\bar i{\rm{B}}}=K_{\bar i}a_k|{h_{\rm{B}}}{|^2}$ and $c_{\bar i 1}=K_{\bar i}K_1, \bar i\in\{\rm{A}, \rm{B}\}$. Since the second and third order leading principle minors are 0, $\frac{{{\partial ^2}{f_2}(t,{x_{\rm{A}}},{x_{\rm{B}}})}}{{\partial {{(t,{x_{\rm{A}}},{x_{\rm{B}}})}^2}}}$ is negative semidefinite and $f_2(t,x_{\rm{A}},x_{\rm{B}})$ is concave.
Therefore, the optimization problem $\mathbf{P}_5$ is convex.
\ifCLASSOPTIONcaptionsoff
  \newpage
\fi
\bibliographystyle{IEEEtran}
\bibliography{refa}

\end{document}